# Simulation for the Evolution of the Australian Netting Spider PM Eye

**Vernon Williams**

vernvlw81@gmail.com

Abstract

This paper reports on a simulated evolution project, which had the goal of simulating the refractive components of the PM eye of Australian netting spider *diopis subrufus* on a desktop computer. The model for the simulation is the anatomy of the eye described by Blest and Land [Ble & Lan 1977]. The evolution simulation was able to produce hundreds of eyes with equivalent optical qualities to the measured eyes for the phenotpe of the netting spider. These artificially evolved eyes began to occur in the computer simulation between $8X10^6$ and $35X10^6$ cycles after start of the computer code. The computer code develops the eye by randomly varying 13 variables that describe the phenotype of spider eye. Previously a paper entitled *Mathematical Demonstration of Darwinian Theory of Evolution*; arXiv 1006.0480 simulated the evolution for the ctenid spider PM eye, *cupiennius sale*. This paper follows the previous paper, but the netting spider eye is more complex than the ctenid PM eye so the simulated evolution equations are more complex.

## 1. Introduction

This paper follows a previous one of similar content, arXiv 1006.0480 entitled *Mathematical Demonstration of Darwinian Theory of Evolution* that uses simulated evolution to evolve the ctenid spider PM eye, *cupiennius sale*. The previous paper showed that simulated evolution for the phenotype agrees with the Darwin concept of evolution. The present paper has a similar purpose, but concerns the evolution of the Australian netting spider eye, which is more difficult to describe mathematically because the eye is a doublet lens rather than a single element as in the case of the ctenid PM eye previously described in arXiv. Darwin pondered whether his evolution theory and its principle of random choice could explain the development of precise organs like the eye, which require accuracy sometimes to a portion of a light wavelength like 0.000005 parts of an inch (~$1.25e^{-4}$ mm). Darwin, in his *Origin of the Species* [Dar 1859], expressed his thinking as follows:

*To suppose that the eye, with all its inimitable contrivances for adjusting the focus to different distances, for admitting different amounts of light and for the correction of spherical and chromatic aberration, could have been formed by natural selection, seems I freely confess, absurd in the highest possible degree. Yet reason tells me, that if numerous graduations from a perfect and complex eye to one very imperfect and simple, each grade being useful to possessor, can be shown to exist; if further, the eye does vary ever so slightly, and the variations be inherited, which is certainly the case; and if any variation or modification in the organ be ever useful to an animal under changing conditions of life, then the difficulty of believing that a perfect and complex eye could be formed by natural selection, though insuperable by our imagination, can hardly be considered real.*

The netting spider has special traits. For example, it has the largest eye of any land invertebrate (the female of the species has up to a 1.4 mm eye aperture for its forward looking eyes PM eyes); an eye resolution of 1.5°, meaning good eye resolution valueøand the spider is a toolmaker. As a toolmaker, the netting spider weaves a rectangular shaped web, which it drops onto a prey from an upper twig perch then the spider pulls the prey thin web guy wires. Further details with this amazing spider can be found on the web at *rmbr.nus.edu.sg/nis/bulletin2009/2009niM247-255.pdf.*

The reason for this choice of the spider eye for this project has several reasons. Certainly one is the challenge by Darwin [Dar 1859] that if one case could be found where his Theory of Evolution did not apply then the total Theory of Evolution would be in jeopardy. Another reason is that the Darwin quote above specially mentions the eye as an example of evolution. Still a third reason is that the PM eye of a spider is easy eye to evaluate mathematically because spider eye is compose of optical elements can be represented as a thickens, which is discussed in any õIntroductoryö optics textbook.

Other authors have examined evolution. However, their explanations are not mathematical explanations, but are qualitative ones. As Richard Feynmen has said, õPeople who wish to analyze nature without using mathematics must settle for reduced understanding.ö The simulation presented here is mathematical in the sense that optical equations describe the phenotype of the netting spider eye, after iterating millions of computer cycles, develops an eye from randomly generated input variables. The computer cycling operates on a HP desktop computer employing TRUE BASIC code, using optical analysis techniques described by Nussbaum [Nus 1998].

The starting point for evolution simulation is not the beginning or origin of life, but at a time after life has begun and prototype spiders existed [Dar 1859]. It is also important to note, that the mathematics of the eye at any point can be a partial eye. However, a partial eye as quoted by Darwin above is still better than its peers with no eyes. Such a circumstance implies that an animal with a non-completed eye, an eye that has not evolved to its maximum condition, can still see better than its peers with no eyes.

The netting spider PM eye used for this projecthas a doublet lens as reported by Blest and Land [1977]. The netting spider PM eye has an extremely small fónumber, but with excellent resolution. According to Blest and Land [Ble & Lan 1977] this eye has of f/# = .58, which is remarkable as any camera fan will attest. The netting spider is nocturnal so it hunts for prey in dim light; hence the need for a low fónumber and large aperture. The netting spider eye measurements made by Blest and Land [Bles & Lan 1977] are the model for the simulated evolution of the PM eye.

A description of the paper follows: 1. Introductionóó a discussion of the general paper content; 2. Procedures for Eye evolution simulationóó a discussion of the computer processor simulation ; 3. Paraxial Calculations;ó- a discussion of paraxial eye calculations using Nussbaum paraxial optical equations ;4.Off-axis calculations-- a discussion of off-axis eye calculations using Nussbaum optical equations; 5. Results óó a description of the results obtained; 6. Implications of the Results--some implications drawn from this project.

## 2. Procedures for Eye Evolution Simulation

Langton [Lang 1987] explained that there are two intertwining concepts used to define a particular animal: the genotype and the phenotype. The phenotype relates to the physical part of the animal, its hair color, its behavior, its weight. All of these the traits can be seen by a visual examination of the animal. For example, a snake is obviously seen as different then a wolf. In the eye case, the phenotype relates to defining the eye shape and its refraction characteristics. The genotype is the biological recipe that defines animal and is located in every cell. The genotype is commonly known as the animalǿs DNA. Thus, the phenotype and genotype are intimately intertwined within animal. In reality, the genotype defines the phenotype; but if the animal is killed or dies before it procreates the death of the phenotype also removes the genotype from the gene pool. The most fundamental way to evaluate a simulation of the evolution of a spider eye would be to work directly with genotype changes as the generations proceed. However, except for small fragments of information [Ghe2005] the method of how the genotype codes and produces a spider eye is unknown. So in this paper the simulated evolution works with the phenotype because the state-of-the-art does not allow otherwise. (For those readers wanting more scientific and philosophical depth into the topics of Darwin Evolution, phenotype, and genotype please read Kuppers [Kup 1990]).

The netting spider PM eye (henceforth, sometimes referred to as the ǿeyeǿ as far as this paper is concerned) is simulated by a two- part optical program. Both parts (really two subroutines) employ equations developed by Nussbaum [Nus 1998 and Nus 1977]. In both cases, the Nussbaum equations use matrix mathematics for calculation. Thirteen variable values first are calculated in the paraxial equations, the first subroutine. Then these variables are plugged into the second subroutine for the offǿ axis marginal rays. Although this procedure is somewhat cumbersome, it is has some advantage. This procedure is common practice in the beginning phase for many commercial optical designs because the paraxial equations are much simpler to calculate because they are analytic whereas, the offǿaxis optical ray trace subroutine requires an optimization process. (Note: paraxial implies that the optical ray is within 5 degrees of the optical axis.) The offǿaxis optical ray trace uses five rays at five intervals spread across the entrance pupil, AP namely rays at positions 0.16/2AP, 0.35/2AP, 0.50/2AP, 0,65/2AP and 0.85/2AP. Even though only five rays is a small number of rays when compared to commercial optics design; however, only five rays are adequate for this simulation.

Most spider eyes will be exactly like their parents. However, when a mutant spider eye does occur, this spider must compete for food and reproduction rights with its siblings and its parents that have do not have mutated eyes.(Atmar [Atm 1997] gives a short set summary of requirements for Darwin evolution .) If the mutant spider eye is superior, the spider with the mutated eye will have an advantage over its peers without this mutation. In this case, Darwin natural selection takes place (survival of the fittest.) In this paper, the simulated natural selection the completed mutated spider eye will have an extremely small f-number coupled with excellent resolution. If the mutant does not have these superior characteristics, it is removed from the gene pool (removed from the code) because it will not be able to compete for food. When the spider is removed from the gene pool it can no longer reproduce and has no progeny. Further discussions on these requirements for a õgoodö eyeö are discussed in Section 4.

According to Darwin, mating is a lesser form of evolution and is not as significant as natural selection [Dar 1859, p 136]. In addition, Mendel [Men 1886] showed that offspring inherit traits from each parent by a probability function that is difficult to specify except through statistics, and is difficult to evaluate in a multi-cycle condition like needed here. Because of the difficulty of unraveling the effects of mating in the simulation, mating is not used. The female netting spider has the largest eye [Ble & Lan ] of the two genders for the netting spider. This paper involves only one gender always spiders, the female. Animal eyes that have õgoodö eye resolution have some method in the refractive part of the eye to correct for spherical aberration. One method for correcting for spherical aberration is to curve the eye lens by adding conics to the spherical surface so that spherical aberration is minimized. The trilobite, an ancient, sea animal that became extinct before the dinosaurs roamed the earth, and in the latter stages of its existence the trilobite had lens surfaces corrected by conic correction [Cla 1975 & Lev 1992].The shape of the eye lens of the trilobite is a known fact because the trilobites had calcite eyes, which were preserved in the fossil record. Conic corrections are the simplest method of correcting the lens surface and are used here. The optical equations developed by Nussbaum [Nus 1998 & Nus 1977] employ this method.

(INTENTIONALY LEFT BLANK)

|  | Radius [mm] | Numeric |
|---|---|---|
| First Surface, R1 | .660 | N/A |
| Second Surface, R2 | .600 | N/A |
| Third Surface,R3 | .490 | N/A |
| Aperture | $\cong 1.40$ | N/A |
| F/Number |  | $\cong .58$ |
|  |  |  |

**Table 1. Blest and Land Measurements for PM Eye of the Netting Spider [Ble & Lan 1977]**

Table1 shows the measurements made by Blest and Land for the PM eye of the netting spider. These measurements represent both science and art because of the difficulty making the measurement. The simulation evolves eyes that have very similar to those of Table 1.

In the simulation, the starting value for three radii of curvature values is a random number between 0 and 1; thus the outputs result varies from cycle to cycle. (One hundred õgoodö eyes is called a õset.ö). The simulation code causes results from set to set to be slightly different because of the use of random numbers in the computer code.

Below is a sample of the TRUE BASIC code that varies X , which in turn changes the variables randomly. The various subprograms listed and called in this sample code also contain random numbers that further define the amplitude and direction of change.

****************_PART OF TRUE BASIC RANDOMIZATION CODE_********************

IF X>0.000 AND X<=.0800 THEN   CALL RMAP(Y,X,C,Z,AP)  ! AP HIT

IF X>.0800 AND X<=.1600 THEN   CALL RMN2(Y,X,C,Z,N2)  ! N2 HIT

IF X>.1600 AND X<=.2400 THEN   CALL RMN3(Y,X,C,Z,N3)  ! N3 HIT

IF X>.2400 AND X>=.3200 THEN   CALL RMN4(Y,X,C,Z,N4)  ! N4 HIT

IF X>.3200 AND X<=.4000 THEN   CALL RMR1(Y,X,C,Z,R1)  ! R1 HIT

IF X>.4000 AND X<=.4800 THEN   CALL RMR2(Y,X,C,Z,R2)  ! R2 HIT

IF X>.4800 AND X<=.5600 THEN   CALL RMR3(Y,X,C,Z,R3)  ! R3 HIT

IF X>.5600 AND X<=.6400 THEN   CALL RMM1(Y,X,C,Z,M1)  ! M1 HIT

IF X>.6400 AND X<=.7200 THEN   CALL RMM2(Y,X,C,Z,M2)  ! M2 HIT

IF X>.7200 AND X<=.8000 THEN   CALL RMM3(Y,X,C,Z,M3)  ! M3 HIT

IF X>.8000 AND X<=.8800 THEN   CALL RMT2(Y,X,C,Z,T2)  ! T2 HIT

IF X>.8800 AND X<=.9600 THEN   CALL RMT3(Y,X,C,Z,T3)  ! T3 HIT

IF X>.9600 AND X<=.9800 THEN LET YMAX=-10  ! ALLOWS RANDOM JUMPS
IF X>.9800 AND X<=1.00  THEN LET YMIN=10  ! ALLOWS RANDOM JUMPS

****************_END OF TRUE BASIC RANDOMIZATION CODE_********************

The randomization process depends on the computer's random number generator. It is not impossible for the computer to generate a true random number [Cha1995]. However, it is possible to generate a random number that is sufficiently random for the purpose here. In order to check whether the computer did generate numbers of sufficient randomness, a test using a billion cycles to judge the applicability for the computer to generate a random number accurate to one part in 10,000, which is sufficient for this investigation. Table 2 shows how the paraxial parameters that very as a function of random choice. Note that the values for N4 are about 7.5 times greater than other variables.

This problem seems to be located in TRUE BASIC itself. However, this difficulty, however, does not cause incorrectness in values of N4.

## 2. Paraxial Calculations

The simulation of the netting spider eye is made in two steps. The first step calculates the variables for the refraction of the netting spider lens for the paraxial condition. The paraxial condition is linear and thus analytic. When a light ray is within 5 degrees of the optical axis, the paraxial conditions are met. The second step is to use the variables calculated paraxially, and then input these variables into the off-axis calculation. The off-axis calculation is accurate across the aperture of the optical input field. However, the off-axis calculation is accurate for about 80 degrees of the input field. Actually, the off-axis calculation could be a one-step method except the computer time would be too long. Using the paraxial calculation first is the usual procedure for commercial optics design.

Because the netting spider has a doublet rather than a singlet eye lens, a complex matrix calculation will result. The solution was determined via the symbolic mathematical software Derive *6* sold by Texas Instruments. Equations 6, 7, 8, 9 show these results, for the Gaussian components of the 2 X 2 matrix result. The Nussbaum optical equations [Nus 1998] are the source for the paraxial calculations. The Nussbaum paraxial calculations generate the eye's focal plane position from the optical variables for each computer calculation cycle. The result of the paraxial calculation is ultimately a 2 X 2 Gaussian optical matrix, described by Nusbaum [Nus 1998]. From this optical matrix result in numerical values for the variables R1, R2, R3, N2, N3, N4, T1, T2, T3, and BFL. These variables are diagrammed in Figure 2. In addition, to these variables list are the conic constants, M1, M2, and M3, used to correct spherical aberration for the spherical lens surfaces. The conic constants are also determined during the paraxial calculations, primarily as a convenience. The M1, M2, M3 variables are not used until the off-axis calculation occurs.

Figure 1 is a diagram of the doublet eye lens of the netting spider as described by Blest and Lamb[Ble &Lan 1977]. Figure 2 is the diagram of effective lens generated from the data of Figure1. The systems equation is taken directly from Nussbaum [Nus 1998, p 20] and is the matrix multiplication **M31=R1 T32 R2T21R1.**This matrix equation can be solved algebraically to obtain the Gaussian 2 x 2 matrix.This matrix equation expresses the doublet shown in Figure 1.

$$ALPHA=T2/N2 \tag{1}$$

$$BETA=T3/N3 \tag{2}$$

$$K1=(N2-N1)/R1 \tag{3}$$

$$K2=(N3óN2)/R2 \tag{4}$$

$$K3=(N4-N3)/R3 \tag{5}$$

With the aid of the mathematical program, Derive 6, made by Texas Instruments, the matrix equation M32 was solved algebraically and is listed below as a function of the Gaussian constants, A, B, C, D.

$$A=óK3*(BETA*K2*(ALPHA*K1ó1)ó(ALPHA+BETA)*K1+1)$$
$$+K2*(ALPHA*K1ó1)óK1 \tag{6}$$

$$B=K3*(ALPHA*BETA*K2óALPHAóBETA)óALPHA*K2+1 \tag{7}$$

$$C=K2*(ALPHA*BETA*K1óBETA)ó(ALPHA+BETA)*K1+1 \tag{8}$$

$$D=óALPHA*BETA*K2+ALPHA+BETA \quad . \tag{9}$$

Other important distances are:

$$FFL=N1*B/-A \qquad \text{(front focal length)} \tag{10}$$

and

$$BFL=N4*C/(óA) \qquad \text{(back focal length)}. \tag{11}$$

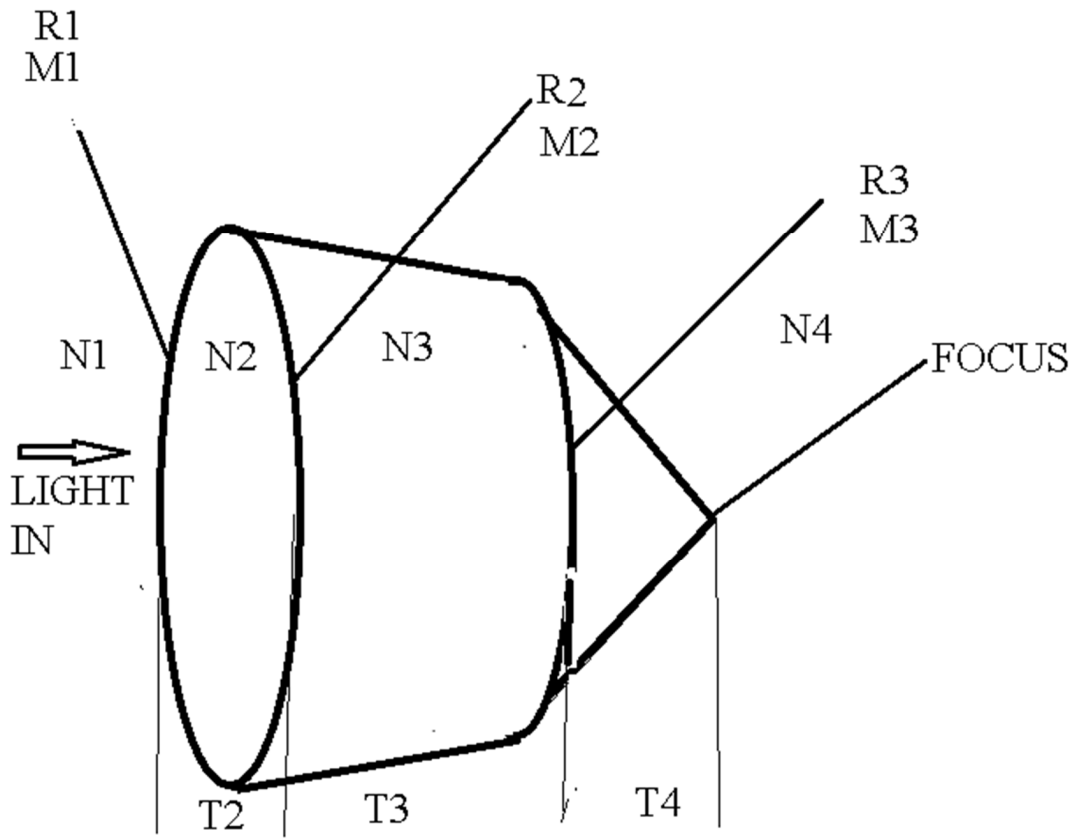

**Figure 1. Netting Spider Eye Side View**

In Figure 1, the surfaces, 1 through 3, are designated by surface 1, radius R1 and conic constant M1, and the same for surfaces 2 and 3. The vertex element thicknesses are represented byT2, T1 through T3. T4 will be represented by BFL shown in Figure2. Notice also the various indexes of refraction, N1, N2, N3, and N4.

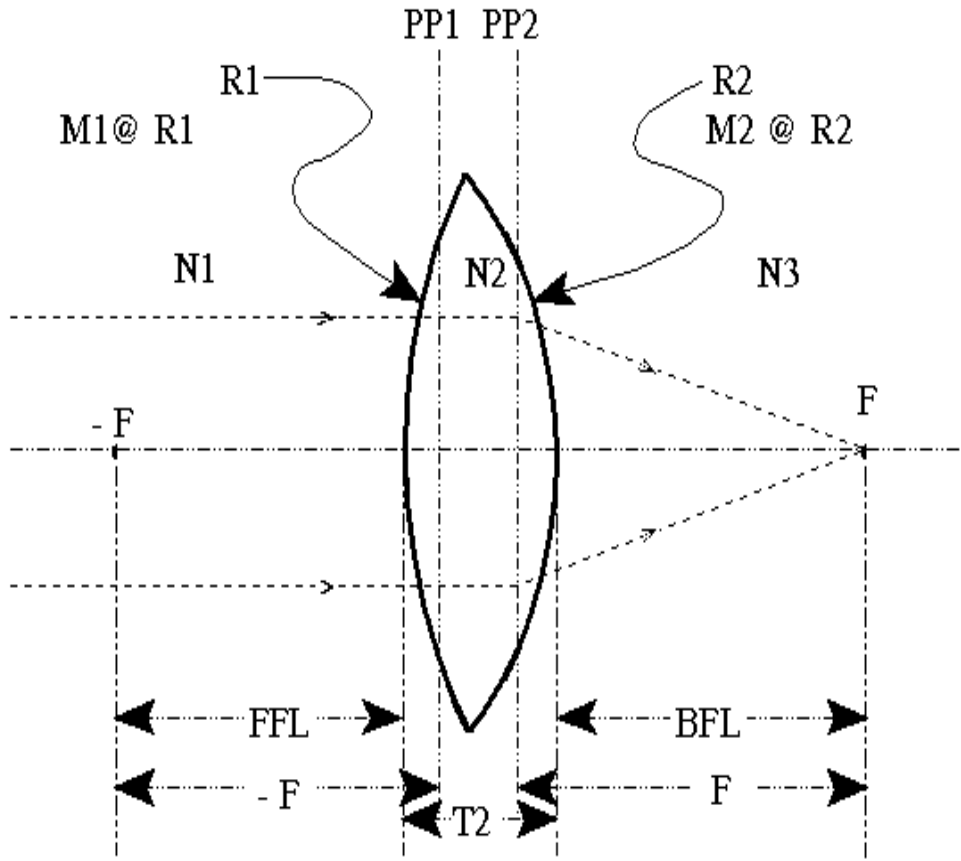

**Figure 2. Effective Singlet Obtained From Netting Spider Doublet**

This effective lens is determined by applying the equations (1) through (11). In effect, this application of the equations converts the lens of Figure 1 into the lens shown in Figure 2.There are things that should be noticed about Figure 2. Light travels from left to right. Note the thickness of the lens, T2, is measured at the vertex of the lens. The focal length of a lens on the left side is - F. On the right side of the lens the focal position is simply F. The spaces where the various indexes of refraction occur are labeled N1, N2, and N3  Also, notice that the Front Focal length, FFL is on the left, whereas, the Back Focal Length, BFL, is on the right of the lens.

| ITEM | #1 | #25 | #50 | #75 | #100 |
|------|------|------|------|------|------|
| R1 | .4123 | .4094 | . 4041 | .4031 | .4006 |
| R2 | ó.5367 | ó.5391 | ó.5441 | ó.5415 | ó.5405 |
| R3 | ó.5133 | ó.5114 | ó.5193 | ó.5209 | ó.5100 |
| T2 | .2824 | .2908 | .2856 | .2796 | .2856 |
| T3 | .7214 | .7262 | .7447 | .7359 | .7359 |
| BFL | .1434 | .1469 | .1382 | .1816 | .1753 |
| N2 | 1.4907 | 1.4829 | 1.4829 | 1.4743 | 1.4775 |
| N3 | 1.4738 | 1.4738 | 1.4842 | 1.5003 | 1.5030 |
| N4 | 1.3600 | 1.3588 | 1.3456 | 1.3550 | 1.3550 |
| AP | 1.3390 | 1.3320 | 1.3485 | 1.3618 | 1.3662 |
| M1 | ó1.8444 | ó.1729 | ó.1755 | ó.1733 | ó.1693 |
| M2 | ó2.7275 | ó2.7391 | ó2.7480 | ó2.7493 | ó2.7556 |
| M3 | ó1.0000 | ó1.0000 | ó1.0000 | ó1.0000 | ó1. 0000 |
| F/# | .5953 | .6088 | .5986 | .6118 | .6080 |

**Table 2. Paraxial Net ting Spider Eye Variables for Sample for 100 "good" Eyes (NOTE: DATA TAKEN FROM SET = B--.02)**

The data from Table 2 is for a typical set of hundred õgoodö eyes with representatives for the whole set of hundred eyes being the eyes in that occur in positions #1, #25, # 50, # 75 and #100.

Because the data in Table 2 uses random numbers to obtain the results, a check on the random generator in TRUE BASIC was made The random test results are shown in Table 3.

| ITEM | 1 million | 1billion |
|------|-----------|----------|
| R1 | 79941 | 80002045 |
| R2 | 80091 | 80006129 |
| R3 | 80234 | 80010331 |
| T2 | 79645 | 79993842 |
| T3 | 79646 | 79993843 |
| N2 | 79948 | 80006312 |
| N3 | 80141 | 79995370 |
| N4 | 679556 | 680001832 |
| AP | 80403 | 79998902 |
| M1 | 79736 | 79989760 |
| M2 | 79926 | 79997594 |
| M3 | 79925 | 79997593 |
| YMIN | 20 196 | 19994050 |
| YMAX | 20039 | 20004822 |

**Table 3. Random Test**

Notice that also, N4 lies between .2400 and .3200 where the extra account occurs for the random numbers, and for some reason this segment in the random generator occurs about 7.5 times the other intervals. However, this result is not an impediment for the total calculation for paraxial variables.

## 4.0 Off–Axis Calculations

After completing the paraxial calculations, that is, all the variables have been determined; the next procedure is the off-axis calculations. The purpose of the off-axis calculations is to make the spider eye acuity and f-number equally valid, or within specification, all across the aperture. Five light rays are used for this purpose, namely at **0.16/2AP, 0.35/2AP, 0.50/2AP, 0.65/2AP and 0.85/2AP.** The equations for doing this are the Nussbaum off-axis optical ray trace program [Nus 1979]. The original purpose was to allow optical ray traces be made on a small calculator. These equations developed by Nussbaum have been adapted for use on a desktop computer used here.

A computer cycle first calculates the paraxial variable values, which are then input it into the off- axis program. However, three of the variables were calculated in the paraxial part are not used until the off axis calculations are made. These variables are the conic variables M1, M2, and M3. An explanation of how the conic variables are used follows for $M_n$:

Mn= -1 implies a spherical conic correction surface

Mn= 0 implies a parabolic conic correction surface

Mn> 0 implies elliptical conic correction surface

Mn <0 implies a hyperbolic corrections

These corrections are added to the value of $R_n$ to complete the off-axis calculations for the surface radii. An important concept to remember is that the final eye values for paraxial variables and off-axis variables are tied together by using the focus location at BFL. This technique ensures that the off axis rays are located at the same focus as paraxial variables.

There are two criteria for good eye in this simulation based upon the data from Blest and Land [Ble & Lan 1977]. The f/# should be about f/# about 0 .58 and the imaged spot size is about 0.02 mm. The True Basic specification to reflect Blest and Land's measurement are coded as f/# must be 0.55 to 0.62 with the acuity specification coded as spot size, SP, is less than 0.02 mm.

## 5 Results

| Item | #1 | #25 | #50 | #75 | #100 |
|------|------|------|------|------|------|
| R1 | .4123 | .4094 | . .4041 | .4031 | .4006 |
| R2 | ó.5367 | ó.5391 | ó.5441 | ó.5415 | ó.5405 |
| R3 | ó.5133 | ó.5114 | ó.5193 | ó.5209 | ó.5100 |
| T2 | .2824 | .2908 | .2856 | .2796 | .2856 |
| T3 | .7214 | .7262 | .7447 | .7359 | .7359 |
| BFL | .1434 | .1469 | .1382 | .1816 | .1753 |
| N2 | 1.4907 | 1.4829 | 1.4829 | 1.4743 | 1.4775 |
| N3 | 1.4738 | 1.4738 | 1.4842 | 1.5003 | 1.5030 |
| N4 | 1.3600 | 1.3588 | 1.3456 | 1.3550 | 1.3550 |
| AP | 1.3390 | 1.3320 | 1.3485 | 1.3618 | 1.3662 |
| M1 | ó1.8444 | ó.1729 | ó.1755 | ó.1733 | ó.1693 |
| M2 | ó2.7275 | ó2.7391 | ó2.7480 | ó2.7493 | ó2.7556 |
| M3 | ó1.0000 | ó1.0000 | ó1.0000 | ó1.0000 | ó1. 0000 |
| SPOTRAD | .0176 | .01271 | .01307 | .01389 | .0143 |
| f/# | .5953 | .6088 | .5986 | .6118 | .6080 |
| START TIME | 43985.862 | óóóóó | óóóó | óóóó | óóóó |
| READ TIME | 53919.247 | 53920.307 | 53921.313 | 53922.326 | 53923.333 |
| J-NUM | 8743957 | 8743994 | 8744019 | 8744047 | 8744072 |

**Table 4. Final Summary Table for a Typical Set of 100 Eyes**

When coded simulation values do not meet the specifications for a õgoodö eye, the simulation code begins anew from the beginning. When 100 are simulated, it is called a set.

Table 4 shows the values for one set of 100 eyes. It should be noted that this is not a quick-running program. The program sometimes runs 10 or more hours to produce a set of 100 eyes. Some sets have run for 35 million cycles before the first õgoodö eye occurs.

## 6 IMPLICATION OF RESULTS

- Although changes in the genotype are the inherited traits that are passed to the next generation of spiders, a great deal of information can be obtained by examining the phenotype. The reason for this is that the genotype is imbedded in the phenotype and vice versa. Hundreds of õgoodö eyes evolved randomly for the netting spider PM eyes using the phenotype.

- Because õgoodö eyes developed in clumps, as shown in Table 4, rather than smoothly, the data gives credence to punctuated equilibrium theory [Eld1972].

- Until the biological state óofó the ó art for understanding how a genotype provides the recipe for a phenotype PM eye, it is not possible to simulate the evolution of the netting spider eye directly from the genotype.

- The question of how many õgoodö could develop from the program variables is unknown, however, the number of statistical possibilities of the variables is immense number. The number of perturbations in this universe is of the order of $10^{10}$.

- A spider egg hatches in about two weeks. This implies all the complex parts of spider including the eyes must form in that time. Adleman [Adl 1994] has shown the rapidity of DNA computing. In the case of the human brain, Castagnoli [Cas 2009] has shown that the amount of data that is processed by human brain is faster than the speed could be obtained by an electronic computer. Castagnoli suggested that some type of quantum computation is required. Thus, it would appear that Nature has developed some type of high-speed, parallel computing.

## Acknowledgments

The author wishes to thank Melenie Maule for correcting his many typing errors. Also thanks go to Pat Williams, Lee Williams, and Mary Kerr for their suggestions to improve readability.